# Clinically Feasible Diffusion Reconstruction for Highly-Accelerated Cardiac Cine MRI

*Shihan Qiu[1,2,3*], Shaoyan Pan[1,4,5*], , Yikang Liu[1], Lin Zhao[1], Jian Xu[6], Qi Liu[6], Terrence Chen[1], Eric Z. Chen[1], Xiao Chen[1], Shanhui Sun[1]*

[1]United Imaging Intelligence, Burlington, Massachusetts, USA
[2]Biomedical Imaging Research Institute, Cedars-Sinai Medical Center, Los Angeles, California, USA
[3]Department of Bioengineering, UCLA, Los Angeles, California, USA
[4]Department of Radiation Oncology and Winship Cancer Institute, Emory University, Atlanta, GA, USA
[5]Department of Biomedical Informatics, Emory University, Atlanta, GA, USA
[6]UIH America, Inc., Houston, TX, USA

**Synopsis** (100-word)
Motivation: The currently limited quality of accelerated cardiac cine reconstruction may potentially be improved by the emerging diffusion models, but the clinically unacceptable long processing time poses a challenge.
Goals: To develop a clinically feasible diffusion-model-based reconstruction pipeline to improve the image quality of cine MRI.
Approach: A multi-in multi-out diffusion enhancement model together with fast inference strategies were developed to be used in conjunction with a reconstruction model.
Results: The diffusion reconstruction reduced spatial and temporal blurring in prospectively undersampled clinical data, as validated by experts' inspection. The 1.5s/video processing time enabled the approach to be applied in clinical scenarios.

**Impact** (40 words)
The proposed diffusion reconstruction pipeline provides a practical solution to cardiac cine reconstruction with enhanced quality for clinical usage. This pipeline may be transferred to the clinical application of other diffusion-based methods.

**Introduction**
Accelerated cardiac cine MRI enables imaging heart motion with shortened or even no breath-holding, thus useful for heart function evaluation in patients. The challenge of a clinical feasible high acceleration centers around the need to reconstruct images from greatly undersampled data within acceptable time. While deep learning (DL) approaches based on convolutional neural networks (CNN) achieved remarkable performance in de-aliasing, their results still suffer from spatial and temporal blurring.[1,2]

Denoising diffusion probabilistic models (DDPM)[3] are an emerging class of generative models, able to provide realistic images. Diffusion-model-based reconstruction have shown potential in improving image quality.[4-9] However, inference with diffusion models relies on an iterative process. For reconstructing 2D multi-coil complex MRI data, minutes to hours of processing time was reported,[4,6-9] which makes diffusion models infeasible in clinical usage. This is more challenging to cine MRI since a video from one slice contains ~20 images, with ~8 slices covering the whole heart.

In this study, we aim to develop a clinically feasible diffusion reconstruction pipeline. A partial diffusion procedure was adopted for a conditional diffusion model, which was constructed to perform enhancement on magnitude data of initial CNN reconstructions, with pseudo data consistency to ensure data fidelity. A multi-in-multi-out processing was designed for dynamic images for speeding up and temporal consistency. With common settings, we achieved diffusion reconstruction with orders of speed-up at an additional 1.5 sec per cine video. The approach underwent evaluation by experts on real-world clinical data.

**Methods**

Diffusion-based enhancement for magnitude MRI

A single-step residual convolutional recurrent neural network (res-CRNN)[1] took the raw k-space and generated initial de-aliased images but with remaining spatial and temporal blurring. A conditional diffusion model input the magnitude image of this DL reconstruction (DLrecon) and made enhancements to it based on the prior knowledge it learned from paired DLrecon and fully sampled images.

Fast diffusion sampling

Applying diffusion models involves a progressive denoising process from Gaussian noises. Instead of the original DDPM method with 1000 steps, an improved DDPM[10] with 50 steps was used. Additionally, during inference, a simulated intermediate step was generated by adding noises to the DLrecon, from which the denoising process started, thereby further shortening the inference to 10 steps.[11]

Multi-in-multi-out model

Considering the 2D+time nature of cine data, a multi-in-multi-out strategy was adopted to improve temporal consistency, in which three consecutive cardiac phases were concatenated as channels. This parallel design also sped up the processing.

Pseudo data consistency

Since res-CRNN reconstruction already enforced data fidelity, a pseudo data consistency (pDC) step was designed on the magnitude images, where the k-space of enhanced images within sampled region was replaced with the ones of DLrecon.

Data and experiments

Cine bSSFP data were collected from volunteers with local IRB approval on 3T MRI scanners (uMR790, UIH, Shanghai, China). **Retro-cine data** of 1071 cine videos from 43 subjects were split into train/validation/test sets with a ratio of 6/2/2. Real-time undersampling masks (acceleration x8~16) were retrospectively applied. Additional **real-time** cine of 16 videos from two subjects were acquired to test the approach in real-world scenarios. Imaging parameters include spatial resolution 1.82x1.82 mm$^2$ and temporal resolution 34ms and 42ms for retro and real-time, respectively. Two experienced (>10yr) experts provided rankings for quality evaluation.

**Results**

As shown in Figure 2, the diffusion model provided images with sharper edges and details as well as less motion blurring than the original DLrecon. Experts' inspection further confirmed this improved sharpness and quality (Table 1). Results in real-time cine were consistent with the retro-cine ones.

The average processing time for one 25-phase slice is 1.5s on a Tesla V100 GPU, which is much faster (>3600) than other diffusion reconstruction methods (Table 2).

**Discussion**

The proposed framework provides high-quality cine images at a time cost of additional 0.06s per image, which is orders faster than existing ones. For sequential multi-slice cine scans, this translates to about an additional 1.5s for the whole stack, making diffusion-based reconstruction applicable in clinical settings.

Images from the diffusion-based method are sharper both spatially and temporally than DLrecon. The diffusion results give cleaner images than the fully-sampled, and are preferred in overall quality given its balanced performance between noise and sharpness.

In this work, the method was implemented in PyTorch with a common setup. In future, the processing time can be further shortened via parallel programming model such as TensorRT.

**Conclusion**

A diffusion reconstruction pipeline was developed that can be translated to clinical usage, as supported by the validated improved image quality and a 1.5s/slice fast inference speed.

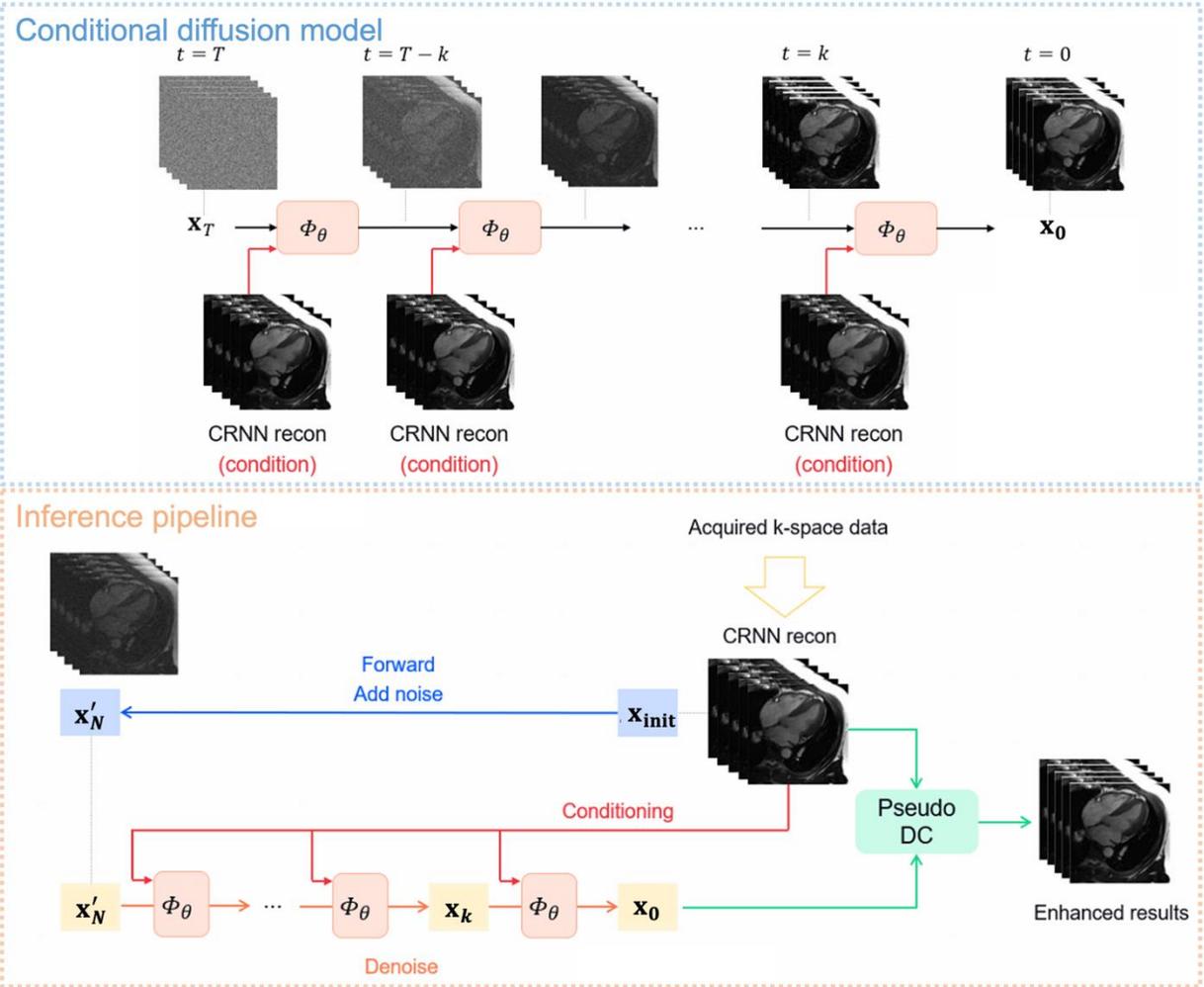

Figure 1. The conditional diffusion model and the inference pipeline. The network was trained to generate high-quality cine data given initial reconstructions. During inference, the k-space is first reconstructed by a res-CRNN, generating coil-combined magnitude images. Then noises are added to simulate an intermediate diffusion step, from which the diffusion model is applied for generation. A pseudo DC step is added at the end to enforce data fidelity before getting the final enhanced reconstruction.

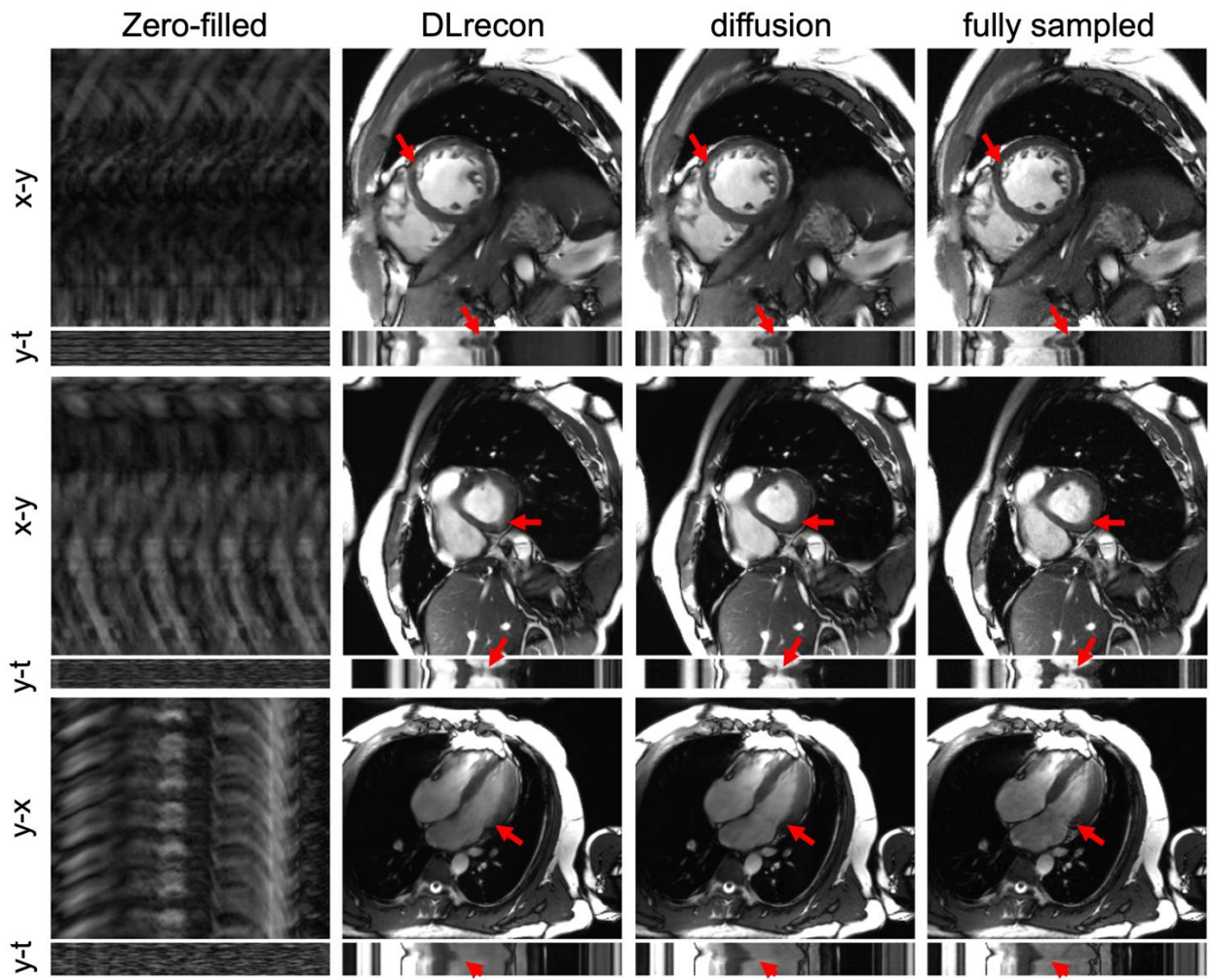

Figure 2. Results on retro-cine data, with fully sampled images as references. The diffusion model showed reduced spatial and temporal blurring than DLrecon, providing better delineation of tissue boundaries.

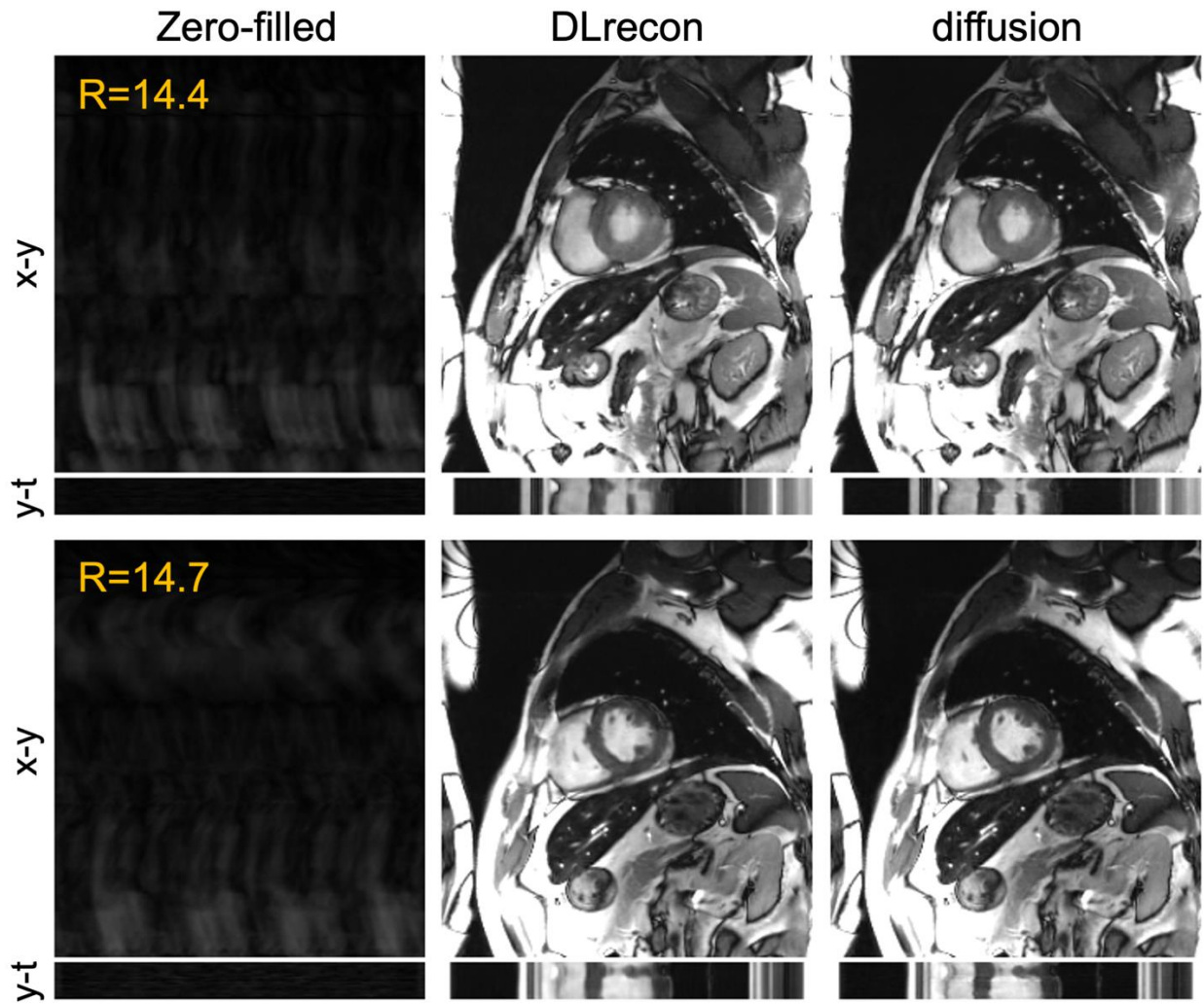

Figure 3. Results on two clinically acquired real-time cine slices with acceleration rates > 14. The undersampling was prospective, therefore with no references available. The proposed diffusion method improved the sharpness at tissue boundaries, mitigated remaining aliasing artifacts, and reduced temporal blurring compared to the original DLrecon.

Table 1. Rankings by two experts on 24 slices of retro-cine and 10 slices of real-time cine. Rank 1 is the best. Bold numbers indicate the best performance among all methods. The asterisks indicate statistically significant differences (P<0.05) compared with DLrecon. The diffusion model achieved better rankings than DLrecon in all aspects, except for SNR in retro-cine data, which resulted from the fact that DLrecon was over-blurry.

|  |  | DLrecon | diffusion | fully sampled |
|---|---|---|---|---|
| Retro-cine | Sharpness | 3.0±0.0 | 1.9±0.2* | **1.1±0.2** |
|  | SNR | **1.1±0.3** | 2.0±0.2* | 3.0±0.1 |
|  | Temporal Sharpness | 3.0±0.0 | 2.0±0.2* | **1.0±0.2** |
|  | Overall Quality | 2.1±1.0 | **1.8±0.4** | 2.1±0.9 |
| Real-time cine | Sharpness | 2.0±0.0 | **1.0±0.0*** | -- |
|  | SNR | 1.9±0.4 | **1.1±0.4*** | -- |
|  | Temporal Sharpness | 2.0±0.0 | **1.0±0.0*** | -- |
|  | Overall Quality | 2.0±0.0 | **1.0±0.0*** | -- |

Table 2. Comparisons of inference time with existing diffusion reconstruction methods[4,6-9]. Our diffusion model achieved an average inference time of 1.5s for one cine video (25 images). Along with the 1s initial reconstruction, the pipeline is much faster than the others, which requires minutes or even hours for one single 2D image.

| Method | Single-/multi-coil | Dimension | Inference time |
|---|---|---|---|
| JaJal, 2021 | multi-coil | 2D | 16 min |
| Chung, 2022 | multi-coil | 2D | 5 hr |
| Chung, 2022 | multi-coil | 2D | 6 min |
| Peng, 2022 | single-coil | 2D | 20 sec |
| Luo, 2022 | multi-coil | 2D | 10 min |
| Gungor, 2023 | single-coil | 2D | 2.2 min |
| **Ours** | **multi-coil** | **2D+time** | **0.06 (+0.04) sec/image, 1.5 (+1.0) sec/video** |